\documentclass[11pt,tightenlines, prd,aps,amssymb,amsmath]{article}
\usepackage[english]{babel}

\usepackage[
a4paper,top=2cm,bottom=2cm,left=3cm,right=3cm,marginparwidth=1.75cm]{geometry}

\usepackage{amsmath}
\usepackage{graphicx}
\usepackage[colorlinks=true, allcolors=blue]{hyperref}

\begin{document}


\begin{flushright}
(\today ) \\
\end{flushright}
\begin{center}
{\Large\textbf{{NFTs: The game is afoot 
 }}} \\[0.5cm]
{\large BK Meister$^{1}$ \& HCW Price$^{2}$}
\end{center}

\begin{center}
$^{1}$FastEagle Holdings \\
$^{2}$Centre for Complexity Science \& Physics Department, Imperial College, 
London SW7 2AZ, UK\\
\end{center}



  \begin{abstract}
  \noindent
On the blockchain, NFT games have risen in popularity, spawning new types of digital assets. We present a simplified version of well-known NFT games, followed by a discussion of issues influencing the structure and stability of generic games. Where applicable, ideas from quantitative finance are incorporated, suggesting various design constraints. Following that, we explain three distinct methods for extracting value from NFT games. The first is to utilise NFT tokens as collateral outside of the game's walled garden; the second is to construct mutual beneficial games based on the participants' risk tolerance, and the third is to use Siegel's paradox in the case of multiple num\'eraires.
  \end{abstract}
  
\section{Introduction: The Games NFT Players Play}
  Non-Fungible Tokens (NFTs) are integral to distributed ledger technology and link unique identifiers on the blockchain to electronic records (and physical items/collectibles).   NFTs gained immense popularity in 2021, after experimentation going back to 2014, with many projects linked to multiple blockchains, with applications in the metaverse, Gamify and the art/Collectibles space. Millions of NFTs have been issued and transferred to the point of congesting the Ethereum network itself\footnote{The Inside Story of the CryptoKitties Congestion Crisis: \url{https://consensys.net/blog/news/the-inside-story-of-the-cryptokitties-congestion-crisis/}}. Market capitalisation and weekly trading volume of NFTs reached billions. For reviews of their development until the middle of 2021, see \cite{Wang2021,Nadini2021}. In 2022,  the market has contracted, and use cases are being reevaluated. A burgeoning niche is NFT games: the topic of the paper where are NFTs integrated into gaming application. A forerunner in this area is `CryptoKitties'\footnote{The website of `CryptoKitties': \url{https://www.cryptokitties.co}.} launched in 2017 and exhaustively described in two papers\cite{Jiang2021, Serada2021}. Over the next several sections the design of games involving NFTs with an emphasis on one of the market leaders, Axie Infinity\footnote{Some circumspect information can be found on \url{https://whitepaper.axieinfinity.com/}, where a fulsome `welcome to our [sic] revolution' is extended.}\cite{Fowler2021,vidal,blog},  will be analysed. Various pitfalls and opportunities will be explored. Questions about advantageous features games should contain are addressed. Particularly appealing might be a mechanism based on the `El Farol bar problem' sometimes called a `minority game'.

Johan Huizinga  in `Homo Ludens'\cite{ludens}    presents games as a necessary condition for culture\footnote{Others have suggested the same about alcohol, smoke, as well as fire. These claims should be viewed with equal skepticism.}. For NFTs,  games are stepping stones towards the gamification as well as the economisation of interactions. This has benefits, since it 
allows the application of well-known quantitative  techniques known from finance, but  it   reduces possibly  the carefree wildness beneficial for learning. 

Virtual games should reflect the variety of real life and   allow multiple players to both compete and cooperate. Winning and as a flip-side losing has   a role. There has to be randomness besides predictability. The existence of  a variety of num\'eraires and risk aversions 
in the player population needs to be incorporated. In addition, the games require a link to the external world.
These requirements 
allow mutually beneficial game designs to exist, provide opportunities, and will be explored in the later part of the paper.
 In the next section simplified definitions are introduced valid for many existing games. 
\section{Definition of Terms in NFT Games}
\noindent
We start by defining in a stylized way properties and terms relevant for NFT Games. In the case
of Axie Infinity and competing games, three types of tokens exist:
\begin{enumerate}
    \item Collectibles with often unique Characteristics,
    \item An Activity Token, and
    \item A Market Place Currency and Governance Token. 
\end{enumerate}
These tokens are game intrinsic, whereas tokens like Ether or a stablecoin provide links to the external non-game world and can act as num\'eraires for intrinsic game assets. On exchanges, the marketplace token is deemed exchangeable for Ether or a stablecoin at a well-defined price, ignoring bid-offer spreads and liquidity issues. 

The NFT Collectibles have within limits unique characteristics, which influence their `breeding' and other gaming properties, e.g. their success in `battles'. This general non-fungibility gives them a collector's value and makes it often hard to establish a precise price. This difficulty will be ignored, and instead, a unique price will be associated with each Collectible. Often the lowest-cost Collectibles have the highest liquidity, and relative value measures are used to price the others. The other two game tokens are fungible. 

Players have to contend with two types of uncertainty. First, the market price evaluated in the num\'eraire of the various in-game tokens varies with time. Second, there is uncertainty about the outcome of the various activities: `breeding', `battle' and `adventure'. These two uncertainties are deemed to be independent, and this simplifies the analysis. In addition, we assume at certain stages of the analysis that the change of the spot price of game tokens in the num\'eraire is determined by the forward price, such that the spot price moves towards the forward price; an idealisation for single-time steps and a helpful simplification. It regards the strong linear drift component in the price as dominant, which over the longer term should overwhelm the random fluctuations, for example, in geometric Brownian motion, which scales with the square root of time.

Next, a formula for the total value at a time $t$ of the Collectibles sector
$$\Phi = \sum^M_{j=0} \hat{a}_j*p_j(t), $$ 
\noindent where $\Phi$ is the amount of capital deployed in the non-fungible Collectibles pool, $\hat{a}_j$ the number of tokens of type $j$ and $p_j$ is the price for a given Collectible in the num\'eraire.  The number of $j$'s can be one, if the token type is unique, or any positive natural number. The number $a_j$ can change in time due to breeding. The associated time index is suppressed here and in other equations to avoid clutter.
One can further subdivide NFTs by users, taken to number $N$, and then arrive at a double sum representing again the market capitalisation of the Collectible sector of the NFT game at a time $t$: 
$$\Phi_{Mcap} = \sum^M_{j=0} \sum^N_{k=0} a_{kj}*p_{j}(t), $$
where $a_{kj}$ is the holding  of the $j$th token by the $k$th user, i.e. $\hat{a}_j=\sum^N_{k=0} a_{k j}$. This ownership matrix will be exceedingly sparse, especially if each Collectible has unique characteristics. In this case, each column with an existing Collectible will have at most one non-zero entry.  The $0$th user is defined to be the game's treasury.
In addition, two other  pools of fungible tokens can be defined representing the activity token and the marketplace currency respectively \\
$$\Psi =\sum_{k=1}^N  b_{k} B(t) = R\cdot B(t) ,$$  and
$$\Omega=\sum_{k=1}^N c_k C(t) 
= S\cdot C(t), $$ 
where $b_k$ is the token position in the second token held by the $k$th user, with $R$ the outstanding number and $B$ the price per token, and $c_k$ is the position held by the $k$th user in the  market token of which $S$ are issued and the current traded price is $C$.  Each of these quantities is further indexed by time, since with respect to the num\'eraire the value varies.  Figure 1 is a sketch of a theoretical NFT game ecosystem.
Next we discuss constraints, which add colour to the game. 
\begin{figure}[h]
\centering
\includegraphics[width=\textwidth]{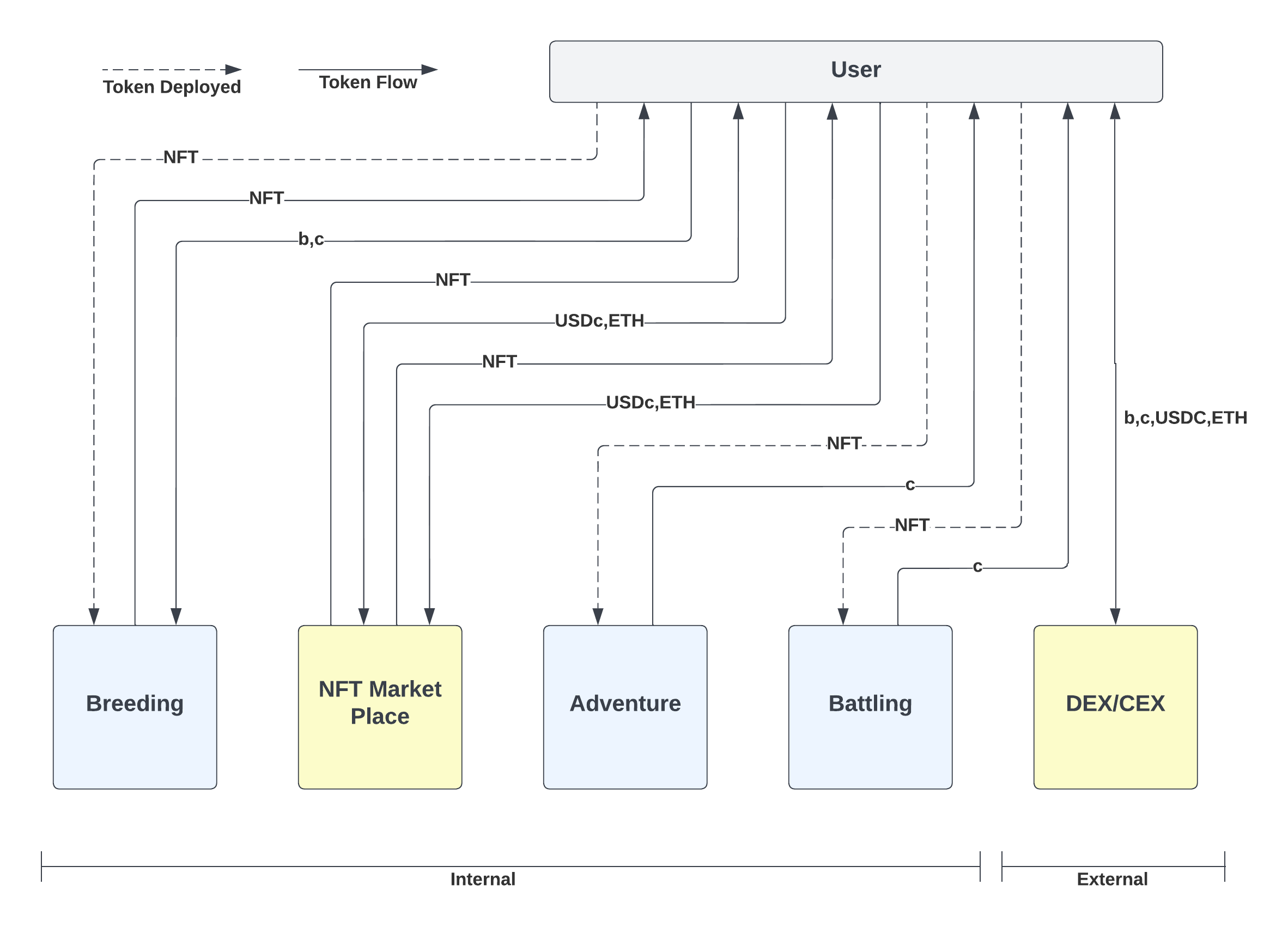}
\caption{\label{fig:frog}NFT game activities}
\end{figure}
\section{Game Constraints and their Implications}
\noindent
In this section, game constraints are described and their  influence on the design of viable games analysed. What will become evident is that some constraints are due to the discrete nature of games, while others are due to possible run-a-way effects. To participate and take part in the various activities described below, a certain number of game tokens are required. This represents an entry barrier and also can lead to the ejection of players after they suffer a losing streak. Run-a-way effects, both hyperinflationary and deflationary, can have a debilitating effect immediately as  players extrapolate trends and act preemptively.

Both constraints reduce the attractiveness of games. Before these consequences are discussed, the description of some further game relevant quantities.
The total theoretical value of all game assets
is given by $$\Pi = \Phi + \Psi +\Omega,$$ where $\Phi, \Psi\,\,\&\,\,\Omega$ are strictly positive. 
This value is not necessarily realisable  due to exchange and liquidity constraints in non-idealised  secondary markets. 
The $k$th user can choose between three activities. `Breeding' and `adventuring', strange as it may seem, are both solitary activities, whereas `battling' often requires multiple users.  

In `breeding', see Figure 1, a player assigns multiple Collectibles to a pool and adds a suitable number of activity tokens that depends on the number of previous births of the lead `breeder'. After a certain number of `births' the `breeding' potential, e.g. seven in the case of Axie Infinity, is exhausted. After a  period of multiple days - 5 days in the case of Axie Infinity - a new Collectible is spawned with characteristics partially inherited and partially randomly chosen. At the same time, the activity tokens are consumed in the process. For simplicity, we assume that breeding takes just one time step to complete and is considered a form of interest rate. 

The randomness in breeding leads to a probability distribution of outcomes. The average outcome for game intrinsic uncertainty is given by the expectation value $\big\langle \cdot\big\rangle_B$. Game risk is idiosyncratic risk, uncorrelated with the rest of the portfolio risk of investors, and unlike systematic risk is diversified out in large enough portfolios.

For the rest of the section the outcome uncertainty involved in `breeding', due to randomness in characteristics of the new Collectible, `adventuring' and `battling' is radically simplified. The probability distribution of outcomes is in each case replaced by the average outcome in value.

The expectation $\big\langle \cdot\big\rangle_P$ associated with the price uncertainty is ignored in the rest of this section\footnote{This contrasts with standard mathematical finance, where risk aversion requires normally a suitable discounting. 

For price uncertainty evaluation, one needs to change into the risk neutral measure or use the pricing kernel approach. In each case, a money market account or short-rate is needed.} and replaced by a unique price, which is either the earlier price or, as an alternative, the forward price.

In all subsequent calculation of gas fees and the interest rates of the num\'eraire tokens are also neglected.

Activity one involves deploying NFT Collectibles to `breed' suitably enhanced by activity ($c$) and market tokens ($b$), this can be done once per time period from $t$ to $t+1$ and a new Collectible, $\hat{a}_{new}$, is generated in the process. For a generic user, $k$ this means that the portfolio value in the deterministic case, 
if activity one is attempted, is 
$$ \bigg\langle \alpha\Big(a_{kj},p_j(t+1),b_k(t+1), c_k(t+1) \Big)\bigg\rangle_B 
=  \sum^{d_1}_{s} a_{kj_s} *p_{j_s}(t ) + p_{new}(t ) -  n\cdot b_k(t+1) -  m\cdot c_k(t+1), $$ where for $d_1 = 2$, i.e. for a parent-child model with $j_1 \,\,\&\,\, j_2$ referring to Collectibles owned by the $k$th users, which are deployed in `breeding'. $n$ and $m$ are the fractions of the activity and market tokens remaining after the cost for 'breeding' has been subtracted. 
In general, the number of tokens required for breeding can vary, and in Figure 1 the relationship between all tokens is shown.

To simplify the model the agent expects the prices to be the same as today, although this will be modified in a subsequent analysis, where instead an explicit forward price is calculated to avoid arbitrage. The constant price version leads to 
$$  \bigg\langle\alpha\Big(a_{kj},p_{j}(t+1),b_k(t+1),c_k(t+1)\Big)
\bigg\rangle_B   \geq  \sum^{d_1}_{s} a_{kj_s}*p_{j_s}(t ) + p_{floor}(t), $$  where the `${floor}$' is the lowest possible Collectible's price. Similar inequalities exist for the other two activities as well.
The assumption to use this minimal price goes beyond using the average and is equivalent to a maximal risk averse investor, who consistently assumes the worst possible outcome. The floor price is pervasive in the NFT space, providing a general index price of any NFT collection, the lowest price tokens are often traded more frequently than highly Collectible ones, leading to more consistent price. 
If the `breeding' is profitable, it will continue unabated until  the price of new Collectibles is driven down far enough to match the cost of the activity tokens required to facilitate the process. 
In other words, to prevent arbitrage, the implied interest rate associated with `breeding' has to match the change in forward price after taking the cost into account. 

The maximal number of Collectible tokens at each time step is given by a modified version of the Fibonacci sequence, where instead of adding the previous two numbers to get the current total, one takes account of all the eligible `breeders'.
`Breeding' restrictions exist, e.g. Collectibles in Axie cannot be bred with their children or siblings along with a breeding limit, causing the numbers of these Collectibles not to rise exponentially. However, limits are game specific but are ignored in the analysis for simplicity.
Such limits force participants to exchange Collectibles to enable the breeding process or possess a large enough supply, since a large enough set of Collectibles through clever mixing can always avoid having to limit its activity\footnote{The exact size depends on the specifics of the restrictions. Anthropology has studied similar elaborate  precautions in various tribes to avoid inbreeding.}. 

Assuming `breeding' is fast, the interest rate of the num\'eraire token appropriately small and the price of the fungible tokens stable enough, then the growth rate of the Collectibles fixes their forward prices, which are assumed to all move in line, 
$$p_j(t+1)  =  \frac{d}{d+1}  p_j(t) + \frac{n\, c(t)+m\, b(t)}{d+1},
$$
where $d$ is the number of input Collectibles needed to breed. If $n\, c /(d+1) +m\, b/(d+1)$ is small, then $d/(d+1)$ describes the dominant term of the gradual price decline.
The growth associated with `breeding' is a natural interest rate, since, ignoring the variability of the Collectibles created, it leads to a steady increase in their number against payment of the `breeding' cycle-specific activity tokens.
If `breeding' is constrained to seven or some other maximal number, then similar to option evaluation backwards from the expiry date on a lattice, one can first evaluate the value of a Collectible with one more `breeding' cycle to go. After this value has been calculated, one adds further and further steps until one reaches the type of Collectible one is considering. The value changes are not identical because the tokens required for `breeding' vary.

The second type of activity is to participate in an `adventure'. This can be written as
$$  \bigg\langle\gamma\Big(a_{kj },  p_{j }(t+1), b_k(t+1)\Big) \bigg\rangle_B =  \sum^{d}_{s=1} a_{kj_s}*p_{j_s}(t+1)  +  n'\cdot b_k(t+1)$$ 
and for $d = 1$  this simplifies to
$$  \bigg\langle\gamma\Big(a_{kj},p_{j }(t+1), b_k(t+1)\Big) \bigg\rangle_B =  a_{kj_s}*p_{j_s}(t+1)  +  n'\cdot b_k(t+1) ,    $$ 
where $n'$ is the fraction (larger than one) of market tokens available after the `adventure'. For short times one can assume the price is the same after the commitment period to get
$$  \bigg\langle \gamma\Big(a_{kj },p_{j }(t+1), b_k(t+1)\Big)\bigg\rangle_B =   \sum^{d}_{s=1} a_{kj_s}*p_{j_s}(t )  +   n'\cdot b_k(t).   $$ 

The third type of activity is  to `battle'. This involves deploying capital and requires other NFT token account holders to participate:
$$  \bigg\langle\beta  \Big(a_{kj },p_{j }(t+1),b_k(t+1)\Big)\bigg\rangle_B =  \sum^{d'}_s  a_{kj_s}*p_{j_s}(t)+  n''\cdot b_k(t), $$ \\
where $d' = 3$ for a three team battle mode and $n''$ is the fraction of market tokens surviving after the battle. Finally, it follows, if an extended Period $P$ is considered and the average is taken to be sufficient, then total earnings for a 
given the number of times $x_1$,$x_2$ and $x_3$ is given by
$$Total ~Earnings = x_1\cdot  \alpha(a,b,c) + x_2\cdot \beta(a,b,c) + x_3\cdot \gamma(a,b,c),$$  
for the activities $\alpha$, $\beta$ and $\gamma$ with the capital costs $a,b \,\,\&\, c$. 
 Given an agent's available actions of $\alpha$, $\beta$ and $\gamma$ with capital costs and total assets available, the resulting average profit and loss can be calculated  as well as the optimal strategy mixture for growth maximising agents.  The probability that an agent is forced to exist  in the game, if a particular strategy is followed, can also be determined. These edge cases related to the discrete nature of the game are not further discussed. 
 
 A high upfront cost for Collectibles or the required activity tokens, presents  a barrier to entry, which in the initial scaling or adoption phase might be highly detrimental and should be avoided. Whereas once players have been enticed to participate, it might be beneficial to raise the proverbial draw-bridge to keep them in. These game theoretic questions are difficult, have many aspects, and will not be further discussed. 

\section{The Glass Bead Game: An Idealised  Game}
\noindent
To develop a quantitative intuition   a simpler, more idealised, NFT game is next presented that again allows three activities  for participants, but has a smaller number of available token types.
The default choice is to stay passive and sit out a round, and as a consequence,  the portfolio remains unchanged.
A simplified version of `breeding' is described next.
\subsection{Breeding to Win}
\noindent
`Breeding' leads to an arbitrage constraint.
At the beginning of the section, unlike above, we describe an equilibrium `breeding' situation, where the exchange rate stays stable during the time considered and no drift in the Collectible token occurs. 
The first  active choice is to `breed' and this increases game tokens by a percentage $C$ but requires a fixed proportion of game-external market tokens, e.g. Ether. It turns a portfolio of $\Big(A,B\Big)$, where $A$ corresponds to the game avatar tokens and $B$ corresponds to the external market token into the portfolio  
$$
\Big((1+C) A,0 \Big),
$$
if one assumes one needs to convert $B$ game avatar tokens, which depends on the exchange rate, to support the `breeding'. 

This then allows a precise formulation of the arbitrage constraint, since
 only $$  A C= B $$ prevents arbitrage. If $$AC>  B,$$ then there is arbitrage in going long `breeding'.
 If $$AC< B,$$ then there is  only an indirectly exploitable arbitrage by going short `breeding'. Direct shorting  is not possible, since `breeding' is not a time-reversible process. 
  
  If the exchange rate is instead taken to be variable, then  to prevent `breeding' to destabilise the game, $C$ has to adjust either internally through the demand curve relating the amount committed for `breeding' to the resulting $C$, i.e. the higher the proportion of outstanding  game tokens committed to  `breeding' the smaller the value of $C$, or externally by linking $C$ to an oracle, which has access to the external exchange rate between the market and the game token. 
  Either method has pitfalls.
 `Breeding' has similarities with the minting process of algorithmic stable coins and suffers from the same challenges. In particular, one has to be aware of Goodhart's law, which states that once a measured index becomes a target it is liable for manipulation and loses its relevance as an independent quantity describing the system.

Idealisations have been employed.  The `breeding' is deterministic, at least some investors are rational, operational risk associated with `breeding' is negligible, and `breeding' is fast compared so that conventional interest rates, i.e. the standard short rate for game extrinsic tokens can be neglected.
The second activity is to participate in a lottery.
\subsection{Lottery as an Adventure}
\noindent
Lotteries hosted by the game organiser can be described, using the terminology from above, as an `adventure'. In the proposed lottery with probability $p$ the player forfeits some committed market tokens and with probability $1-p$  the player gets a combination of additional game and market tokens. Depending on the pay-out, the lottery requires a subsidy, is self-funding or is profitable on average for the sponsor.
This allows 
the calculation of the associated Sharpe ratios. As a consequence, games can be  compared and ranked. This has implications  discussed in a later  section.
The third option is a modified version of a  `battle' called in the literature a `minority game'.
\subsection{Minority Games: `The last will be first, and the first last'}
\noindent
`Minority games' have been discussed  in the literature as a way to induce unusual patterns of cooperation and can be used to enliven `battles' in NFT games. The rules are simple. Participants add their stakes to one of two sides until a cut-off time. The tokens of the majority  are distributed proportionally to the minority. 
As an example, if the total amount  $b$ is awarded to the winning side $a$, i.e. $b>a$, and the $i$-th participant on the winning side then gets $$ \frac{  b}{a} x_i, $$ with $a=\sum_{j\in{winning\_side}} x_j$ and $b=\sum_{j\in{losing\_side}} x_j$. 
The challenge of the game is to balance the incentive for joining the minority side to enhance one's winning stake with the danger of in the process tipping the scales. Aiming to maximize profits entails risk. 

Instead of awarding the stakes to the minority, one can use other criteria, i.e. base the allocation on the outcome of a football game or the flip of a coin. 
Various other modifications are possible.
To induce players to participate, each minority game can have a sponsor adding $S$ to the pot so that the winner takes $\pi (X_1+X_2) + S $, which is loss making for the sponsor unless $\pi (X_1+X_2) \geq S$. The choice of  $\pi$ is critical to determine the attractiveness of the game. As a complicating factor, the sponsor can always add to the pot, which muddies the water.
To avoid an unproductive rush close to the cut-off time of the game, one can randomize the cut-off or turn it into a conditional stopping time. As an example, when a fixed number of tokens are in the pool, then the game ends. The variations to consider and the strategies to be employed are manifold. The four volumes of ``Winning Ways for Your Mathematical Plays''\cite{conway} contain many ingenious suggestions.
Portfolio aspects of games are discussed next.


\section{How to Design the Optimal Game: Comparing Investments and the Portfolio Approach}
\noindent
Within each NFT game, different strategies exist. These can be compared using measures popular in finance. This includes the well-established methodology of Sharpe ratios, which is the  return minus the riskless rate, i.e. the excess return, in the numerator and the volatility in the denominator. It scales with the inverse of the square-root of time, i.e. $1/\sqrt{T}$. The other quantity of interest is the related optimal investment ratio, which determines the optimal allocation for the available pools and   is, in the one dimensional case, the excess rate of return over the square of the volatility.
In the multidimensional case the numerator becomes a vector and the denominator is replaced by the Moore-Penrose inverse of the covariance matrix, i.e.
 $$(\Vec{\mu} - 1\Vec{I})(\sigma^{dagger}\sigma)^+. $$
 with $A^+$ the generalised Moore-Penrose inverse that coincides with the conventional inverse for invertible matrices but extends to any matrix\cite{lv1,lv2}.


As a slightly offbeat aside, some comments about the development of `monopoly', rules evolve in the world of board games, and it takes time to reach an equilibrium state. The game of `monopoly' was invented more than one hundred years ago as a way to promote a `single tax theory’ and to show not unreasonably the dangers of monopolies. It took about one generation of experimentation until the more or less recognisable version of `monopoly' finally emerged. Why did it take so long because there are a number of tuneable parameters? Various fees and benefits define the relationship between the players and the bank, and in the same way, the relationship between the players themselves can be tuned. There is no need to dwell on details, but it is not hard to see that one can easily destabilise the system by, for example, increasing fees dramatically. One can look at the expectation value of the cash transfers between bank and players per turn and thereby calculate the average increase in the money supply. On the internet, some numbers are discussed concerning the necessary inflation to keep the game interesting. Besides an increase in the average money the players hold, one can also look at the average evolution of the GDP due to building activity. Again, without going into details, the game economy is sensitive to these factors, and a steady decrease in the average cash in the hands of the players would terminate the game.
Three special cases that can enhance NFT games are considered next in dedicated subsections, starting with the `two envelopes game'.

\subsection{Two Envelopes Game}
\noindent
Let's examine the case where players have different num\'eraires. Two of these NFT players with distinct num\'eraires decide to play the `envelopes game'\footnote{This is akin to the well-known Siegel's paradox\cite{beef} and a modified version was also  discussed in a paper by Fisher Black\cite{NBlack1989} about the disadvantage of a fully hedging  foreign exchange risk in an investment portfolio.}.  Each puts equal amounts at the current exchange rate denominated in their preferred num\'eraire into  virtual envelopes. These envelopes are exchanged and at the end of the period they are opened. Gains and losses in the exchange rate are due to the Jensen inequality asymmetric. As an example, assume the exchange rate can either double or half. Each outcome happens with probability one half. Then at the end of the period, when both players open their virtual envelopes, one will have doubled their money, while the other has lost half. The average gain of each player will be an eye-watering $25\%$. Naturally, this can   also be achieved, if investors are willing to take for example foreign exchange risk\footnote{The implication for portfolio allocation is explained in Black\cite{NBlack1989}.}. The advantage of using not game extrinsic, but two game intrinsic currencies would be that it is  easier to adjust an internal exchange rate  to enable   enticing returns, while diverting possibly some of the surplus gain to the game promoters.  Next, another way to improve attractiveness of NFT games.
\subsection{NFTs as Collateral}
\noindent
In this section, it is explained how to link the walled garden of an NFT game in a non-obvious way to the outside world and in the process create value for players. 
The in- and out-flows of non-game external tokens has to  match. It is generally agreed that isolated games deter participation, since assets are locked up which otherwise could accumulate interest.  
Nevertheless, there is a seemingly innocent way of creating value out of nothing. 
How is this done?
If  game intrinsic tokens can be used as collateral on lending platforms in the `real' world for loans or other financial transactions, then their virtual value
becomes real. 
This enables `real' money creation. 
Loans as a result become either outright available or at least cheaper. Game players can use them for investment in the `real' world to create excess yield. The paper value otherwise locked up in the game can turn into something as concrete as the ownership of `real' assets. This diffusion of value through the initially impermeable game boundary is maybe the primary way of creating value. 
Only if the collateral needs liquidation, does this put pressure on the value of the game tokens and can in the process undermine the viability of the game. 

One possible side effect of the  access to additional capital is to entice NFT game promoters to use it to push up the value of their game intrinsic tokens. As the value increases additional funds can be borrowed against the collateral that can again flow into the NFT game creating a  virtuous cycle. This process continues until some external events causes an interruption and leads to a dramatic unwind of leveraged positions.
 Next, the influence of heterogeneous risk aversion on games.
\subsection{Heterogeneous Risk Aversion \& Utility: Borges Babylonian Lottery}
\noindent
Games are seductively attractive, if 
the expected utility of all players is raised during play. This seems at first instance impossible, even in a culture where everybody is declared a star and the majority believes itself to be above the median. Nevertheless, it can under special circumstances be achieved. One such example is a population of game players with heterogeneous risk aversion or utility.  
This is not too far-fetched, since depending on their disposition and history,  NFT players are likely to be more or less risk-adverse. 
Some participants might be thrill and risk seeking gamblers\footnote{Unusual risk preferences are discussed in the Babylonian lottery  by the Argentinian writer Jose Luis Borges.} with a penchant for lotteries with high variance and an overall negative drift, while others have a more conventional log or power utility that requires a positive expected reward for any non-diversifiable risk.
These different outlooks allow the combination into games that can provide a win-win situation for all players, while allowing marginal outflow to the game organisers.

To substantiate the claim, a toy model will next be given.
Take a game that pays out to  risk-adverse player on average a small dividend, while the risk-seeker has on average a negative pay-out but in the style of a lottery on rare occasions a 
`life-changing' win. The diverse utility of all is lifted. The mathematics for such a case is simple and given in the form of a worked out example. 

Assume one risk seeking and ten risk averse players join a game. Each of the eleven players puts their life-savings of one token into the pool.
With the high probability of $95\%$ the risk-adverse players get a $5\%$ payout and the risk seeker loses $50\%$. In the complementary $5\%$ of cases the ten investors each lose $10\%$ leading to doubling of the money of the risk seeking investor. The average gain of the risk-adverse gamblers is $4.25\% $, while the risky gamble has an average loss of $37.5\%$ but a substantial, for a thrill seeker life-enhancing, variance.

The result above can be reformulated. Utility of the game players can be lifted through redistribution. A `propitious game' could be defined as a redistribution  of committed  tokens by players   such that  either the average utility increases\footnote{This is discussed in detail for all players having the same utility function in Hughston   {\it et al.} \cite{hughston} based on    thermodynamic arguments.} or each individual utility increases. 
As stated above, if players possess heterogeneous utilities, then it is possible to construct examples such that for all pool participants, their respective utilities increase. Non-standard  utility could also be associated with other aspects of games, e.g. a pleasurable gaming experience.

\section{Conclusion: Games People Play}
\noindent
This paper covered structures and dynamics prominent in NFT games, along with various valuation methods and mechanisms for improving their attractiveness.   
The psychology and motivations of NFT game players are outside the scope of the paper, except as they relate to game design and dynamics. 
The transparency of the blockchain, i.e. the past is an open book,  is a defining feature of NFT games. What are the consequences? Everybody can observe the reputation of players, and  this could also be done by dedicated monitoring services, which could in addition categorise the agent's past behaviour and use it to predict future action in the settings of the various pools. Predictions could lead to curious feedback loops where players adjust their behaviour in the awareness of being observed. Such adjustments lead to an infinite regress that could lead to instabilities as described in various game theory set-ups.\footnote{See the blog entry about the ``The blue-eyed islander's puzzle'' by Terence Tao  (\color{blue}
$https://terrytao.wordpress.com/2011/04/07/the-blue-eyed-islanders-puzzle-repost/$
\color{black}) 
and links mentioned therein.
}. 
Could there be an advantage in anonymous games? In the same vein, can reputation become a transferable asset? These and other questions need careful analysis and will be dealt with elsewhere.
In general, if reputation has value, then consistency in behaviour matters. A truth generally acknowledged is that if the utility from game participation after transaction cost is positive, games gain popularity. This leads to several tactical challenges for sponsors of games. How should subsidies be optimally employed? All at once in the beginning, or should they be dangled like a `carrot'? The same is true for exit barriers. Should this `stick' be large or small? What are the determining factors? What is the consequence of competitors over-promising benefits of their games? These and similar questions will be discussed elsewhere. 
NFT games are at an embryonic stage of development, and the discovery of a viable game design that can stand the test of time still lies in the future.



\end{document}